\documentclass[apj,twocolumn,twocolappendix]{aastex6}

\usepackage[encapsulated]{CJK}
\usepackage{ucs}
\usepackage[utf8x]{inputenc}
\newcommand{\cntext}[1]{\begin{CJK}{UTF8}{gbsn}#1\end{CJK}}

\shorttitle{On the absence of Non-thermal X-ray Emission around runaway O stars}
\shortauthors{Toal\'{a}, Oskinova \& Ignace}

\usepackage{rotating}
\usepackage{color}
\usepackage{graphicx}	
\usepackage{amsmath}	
\usepackage{amssymb}	
\usepackage{bm}		

\begin{document}

\title{On the absence of Non-thermal X-ray Emission around runaway O stars}


\author{J.A.\,Toal\'{a}\,\cntext{(杜宇君)}$^{1}$}
\author{L.M.\,Oskinova$^{2}$}
\author{R.\,Ignace$^{3}$}

\affil{$^{1}$Institute of Astronomy and Astrophysics, Academia Sinica\,(ASIAA),
  Taipei 10617, Taiwan\\
$^{2}$Institute for Physics and Astronomy, University of Potsdam, D-14476 Potsdam, Germany\\
$^{3}$Department of Physics and Astronomy, East Tennessee State University, Johnson City, TN 37614, USA}

\begin{abstract}
  Theoretical models predict that the compressed interstellar medium
  around runaway O stars can produce high-energy non-thermal diffuse
  emission, in particular, non-thermal X-ray and $\gamma$-ray
  emission.  So far, detection of non-thermal X-ray emission was
  claimed for only one runaway star AE~Aur. We present a search for
  non-thermal diffuse X-ray emission from bow shocks using archived
  {\it XMM-Newton} observations for a clean sample of 6
  well-determined runaway O stars. We find that none of these objects
  present diffuse X-ray emission associated to their bow shocks,
  similarly to previous X-ray studies toward $\zeta$\,Oph and
  BD$+$43$^{\circ}$3654. We carefully investigated multi-wavelength
  observations of AE~Aur and could not confirm previous findings of
  non-thermal X-rays. We conclude that so far there is no clear
  evidence of non-thermal extended emission in bow shocks around
  runaway O stars.
\end{abstract}


\maketitle

\section{INTRODUCTION}
\label{sec:intro}

Massive ($M_\mathrm{i} > $~10~M$_{\odot}$), runaway ($v_{\star}
\gtrsim$~30~km~s$^{-1}$) stars are able to produce large-scale bow
shocks in the Interstellar Medium (ISM). These shocks are driven by
the interaction of the fast stellar wind ($v_{\infty} \gtrsim
$~1000~km~s$^{-1}$), large proper motion, and the ISM. The gas and
dust in the pile-up material are heated and ionized by the strong UV
radiation from the star which makes the bow shock observable at
optical and infrared (IR) wavelengths \citep[e.g.,][and references
  therein]{Kaper1997,Peri2015}.

\citet{Benaglia2010} analyzed Very Large Array (VLA) observations of
the runaway O star BD$+$43$^{\circ}$3654 and concluded that the radio
emission is spatially coincident with the bow shock detected in IR
images. More importantly, this extended radio emission was found to
have a non-thermal origin. Benaglia et al. argued that the non-thermal
origin of the radio emission is produced by syncrotron emission. The
electrons that generate this emission could upscatter photons from
stellar and dust radiation fields through the inverse Compton process,
producing high-energy emission. This interesting detection opened a
new window for exploring the production of non-thermal emission around
massive stars, and a number of theoretical works addressing this
phenomenon have been published \citep[e.g.,][and references
  therein]{delValle2015}.

\citet{delValle2012} presented detailed analytical predictions for the
non-thermal emission from bow shocks around O-type stars. These
authors applied their model to the well-know and closest runaway star
$\zeta$\,Oph, concluding that high-energy emission should be detected
toward its bow shock. However, X-ray and $\gamma$-ray emission has
been eluding detection towards known runaway
stars. \citet[]{Schulz2014} presented Fermi $\gamma$-ray Space
Telescope observations of a sample of 27 bow shocks (including
$\zeta$\,Oph) accumulated over 57 months with no positive
detections. \citet{Schulz2016} extended this study up to 73 bow shocks
using the H.E.S.S. telescopes in the TeV regime with the same
conclusions. In X-rays no detections were obtaned either, despite the
dedicated observational campains using {\it Chandra}, {\it
  XMM-Newton}, and {\it Suzaku} X-ray telescopes
\citep[][]{Terada2012,Toala2016}.


\begin{figure*}
\begin{center}
  \includegraphics[angle=0,width=.95\linewidth]{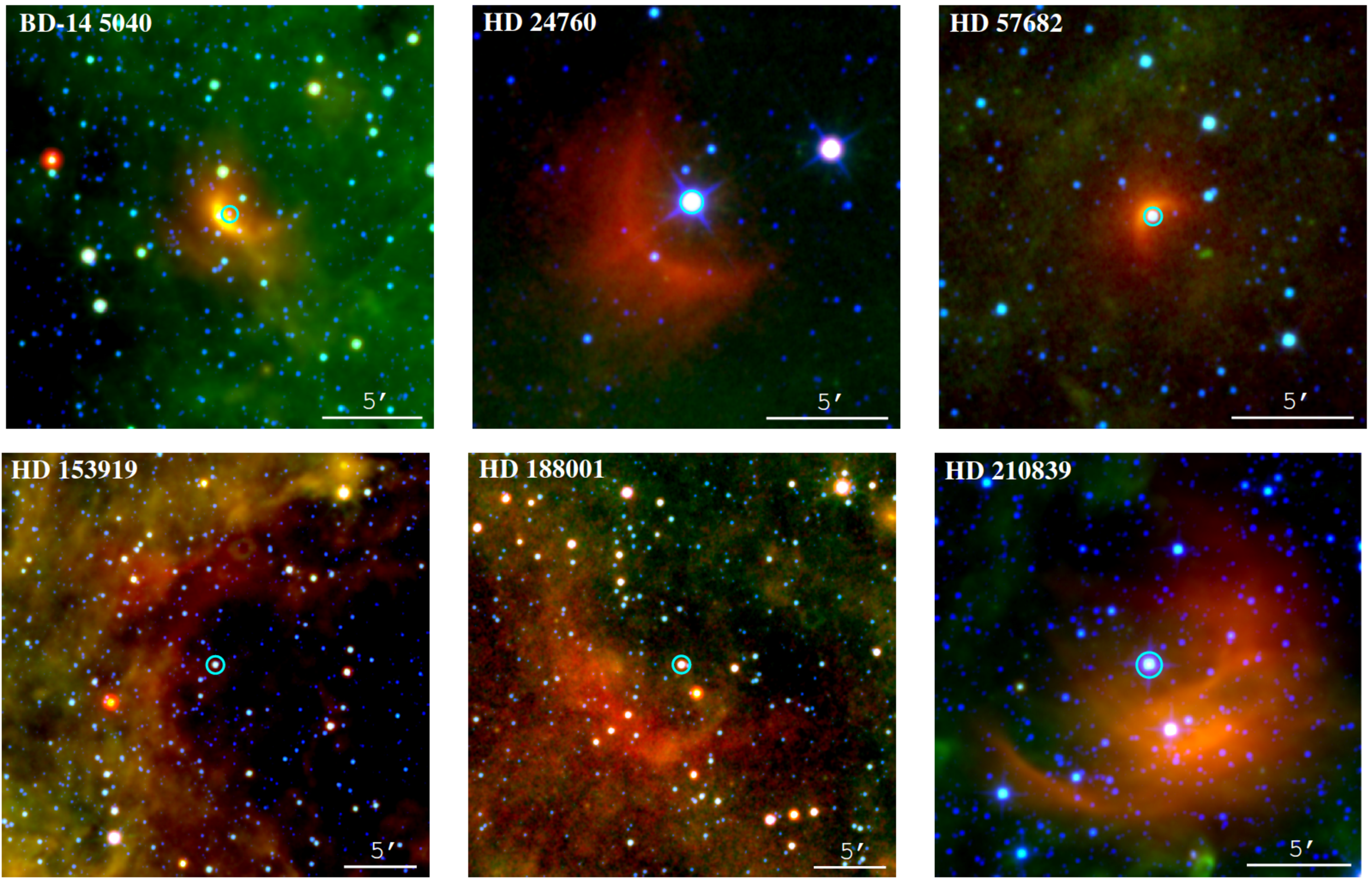}
\label{fig:WISE_RGB}
\caption{Composite colour mid-IR {\it WISE} images of the bow shocks
  around the O stars studied in this paper. Red, green, and blue
  correspond to the W4 (22~$\mu$m), W3 (12~$\mu$m), and W2
  (4.6~$\mu$m) bands, respectively. A circular aperture on each panel
  shows the position of the stars. North is up, east to the left.}
\end{center}
\end{figure*}

There has been only one claim of detection of non-thermal X-ray
emission toward a runaway star. \citet{LopezSantiago2012} presented
{\it XMM-Newton} observations of AE~Aur and reported the discovery of
a ``blob'' of X-ray emission at $\gtrsim$30$\arcsec$ northeast from the
star. These authors also presented a model to explain their results,
but we notice that their spectral analysis cannot be used to
discriminate between a thermal and a non-thermal origin. Furthermore,
these authors compare their {\it XMM-Newton} observations with
low-resolution mid-IR {\it WISE} observations. Under the assumption
that this detection is related to AE~Aur, \citet{Pereira2016}
presented further analytical modelling for this X-ray emission. They
concluded that non-thermal processes in bow shocks around runaway
stars are responsible for a significant fraction of the high-energy
photons produced in our Galaxy.

In this paper we present a search for non-thermal diffuse X-ray
emission in bow shocks around O-type stars. We use archived {\it
  XMM-Newton} observations of a sample of well-determined galactic
runaway stars. Section~2 presents our sample and describe the {\it
  XMM-Newton} observations. In Section~3 and 4 we presents and discuss
our results, respectively. Finally, we present our conclusions in
Section~5.

\section{Observations - The sample}

To obtain a clear sample of runaway O stars we searched the list
presented by \citet{MA2016}. These authors identified runaway stars
using the proper motions reported by the first Gaia Data Release
\citep[DR1;][]{Brown2016}. Their table~1 presents a list of confirmed
candidates as well as a list of new discoveries. We cross-corretated
that table with archived {\it XMM-Newton} EPIC observations and clear
detections of bow shocks in the {\it WISE} W4 22~$\mu$m or {\it
  Spitzer} MIPS 24~$\mu$m band. As a result, our sample consists of 6
objects: BD$-$14$^{\circ}$5040, HD\,24760 ($\epsilon$\,Per),
HD\,57682, HD\,153919, HD\,188001 (9\,Sge), and HD\,210839
($\lambda$\,Cep). Figure~1 presents mid-IR colour-composite {\it WISE}
images of the six objects studied in this paper.

Details of the X-ray observations used in this paper are given in
Table~1. Columns 5, 6, and 7 of Table~1 present the total exposure
time for the pn, MOS1, and MOS2 EPIC cameras, respectively. It is
worth mentioning that almost all observations were performed with deep
exposures ($t_\mathrm{exp} > 30$~ks), except for the cases of
HD\,24760 and HD\,57682 ($t_\mathrm{exp} \gtrsim 10$~ks). Observations
of BD$-$14$^{\circ}$5040 were only performed using the MOS cameras but
with deep exposure times of $t_\mathrm{exp} \sim 70$~ks. Finally, we
remark that the EPIC cameras have a FWHM$\approx$6\arcsec.

\begin{table*}
\caption{Observation details}
\begin{center}
\begin{tabular}{|ccrc|ccc|ccc|}
\hline
\hline
\multicolumn{1}{|c}{Object} & \multicolumn{1}{c}{R.A.} & \multicolumn{1}{c}{Dec.} & \multicolumn{1}{c}{Obs.\,ID.} & \multicolumn{3}{|c}{Total exposure time} & \multicolumn{3}{|c|}{Net exposure time} \\
\multicolumn{1}{|c}{} & \multicolumn{1}{c}{(J2000)} & \multicolumn{1}{c}{(J2000)} & \multicolumn{1}{c}{} & \multicolumn{1}{|c}{pn} & \multicolumn{1}{c}{MOS1} & \multicolumn{1}{c}{MOS2} & \multicolumn{1}{|c}{pn} & \multicolumn{1}{c}{MOS1} & \multicolumn{1}{c|}{MOS2} \\
\multicolumn{1}{|c}{} & \multicolumn{1}{c}{} & \multicolumn{1}{c}{} & \multicolumn{1}{c}{} & \multicolumn{1}{|c}{(ks)} & \multicolumn{1}{c}{(ks)} & \multicolumn{1}{c}{(ks)} & \multicolumn{1}{|c}{(ks)} & \multicolumn{1}{c}{(ks)} & \multicolumn{1}{c|}{(ks)} \\
\hline
BD$-$14$^{\circ}$5040 & 18:25:38.90   & $-$14:45:05.74  & 0742980101 & $-$   & 72.62  & 72.63 & $-$   & 69.52 & 69.55 \\
HD\,24760            & 03:57:51.23   &    40:00:36.78  & 0761090801 & 14.04 & 15.66  & 15.63 &  6.42 &  8.70 & 8.90  \\
HD\,57682            & 07:22:02.06   & $-$08:58:45.77  & 0650320201 & 8.73  & 11.45  & 11.47 &  6.02 &  8.87 & 9.47  \\
HD\,153919           & 17:03:56.88   & $-$37:50:38.91  & 0600950101 & 42.78 & 50.27  & 50.37 & 41.40 & 50.27 & 50.07 \\
HD\,188001           & 19:52:21.76   &    18:40:18.75  & 0743660201 & 26.76 & 31.34  & 31.32 & 26.76 & 31.34 & 31.32 \\
HD\,210839           & 22:11:30.57   &    59:24:52.15  & 0720090501 & 83.82 & 93.94  & 93.99 & 73.49 & 92.70 & 92.34 \\
\hline
\hline
\end{tabular}
\end{center}
\label{tab:multicol}
\end{table*}

\section{Analysis and Results}

The {\it XMM-Newton} EPIC observations were processed using the
Science Analysis Software (SAS) version 15.0 and the calibration
access layer available on 2017 January 6. All observation data files
(ODF) were processed using the SAS tasks {\it epproc} and {\it emproc}
to produce the event files. In order to identify and excise periods of
high background level, we created light curves of the background,
binning the data over 100~s for each of the EPIC camera. The final
net exposure times for each EPIC cameras are listed in columns 8, 9,
and 10 in Table~1.

To unveil the presence of diffuse X-ray emission around the runaway
stars studied here, we made use of the XMM-ESAS tasks which are
optimized for the identification of extended sources, to produce
images in different energy bands. These algorithms also help identify
point-like sources projected in the line of sight of the bow
shocks. For each object, we created exposure-map-corrected,
background-subtracted EPIC images in three different energy bands,
namely 0.3--1.0 (soft), 1.0--2.0 (medium), and 2.0--5.0~keV (hard),
following the Snowden \& Kuntz's cookbook for analysis of {\it
  XMM-Newton} EPIC observations of extended objects and diffuse
background \citep{Snowden2011}. All images have been adaptively
smoothed using the ESAS task {\it adapt} requesting 10 counts under
the smoothing kernel of the original images. The resultant images for
each target are presented in Figure~2.

Figure~2 shows that all progenitor stars are detected in X-rays as
well as a large number of point sources in the field of view of the
observations. Furthermore, Fig.~2 clearly shows the absence of diffuse
X-ray emission in the six bow shocks around our targets. To highlight
the absence of detected X-ray emission from bow shocks, all panels of
Fig.~2 also present contours corresponding to the mid-IR emission as
detected by the {\it WISE} W4 22~$\mu$m band, where the extended X-ray
emission was expected.

\begin{figure*}
\begin{center}
  \includegraphics[angle=0,width=.95\linewidth]{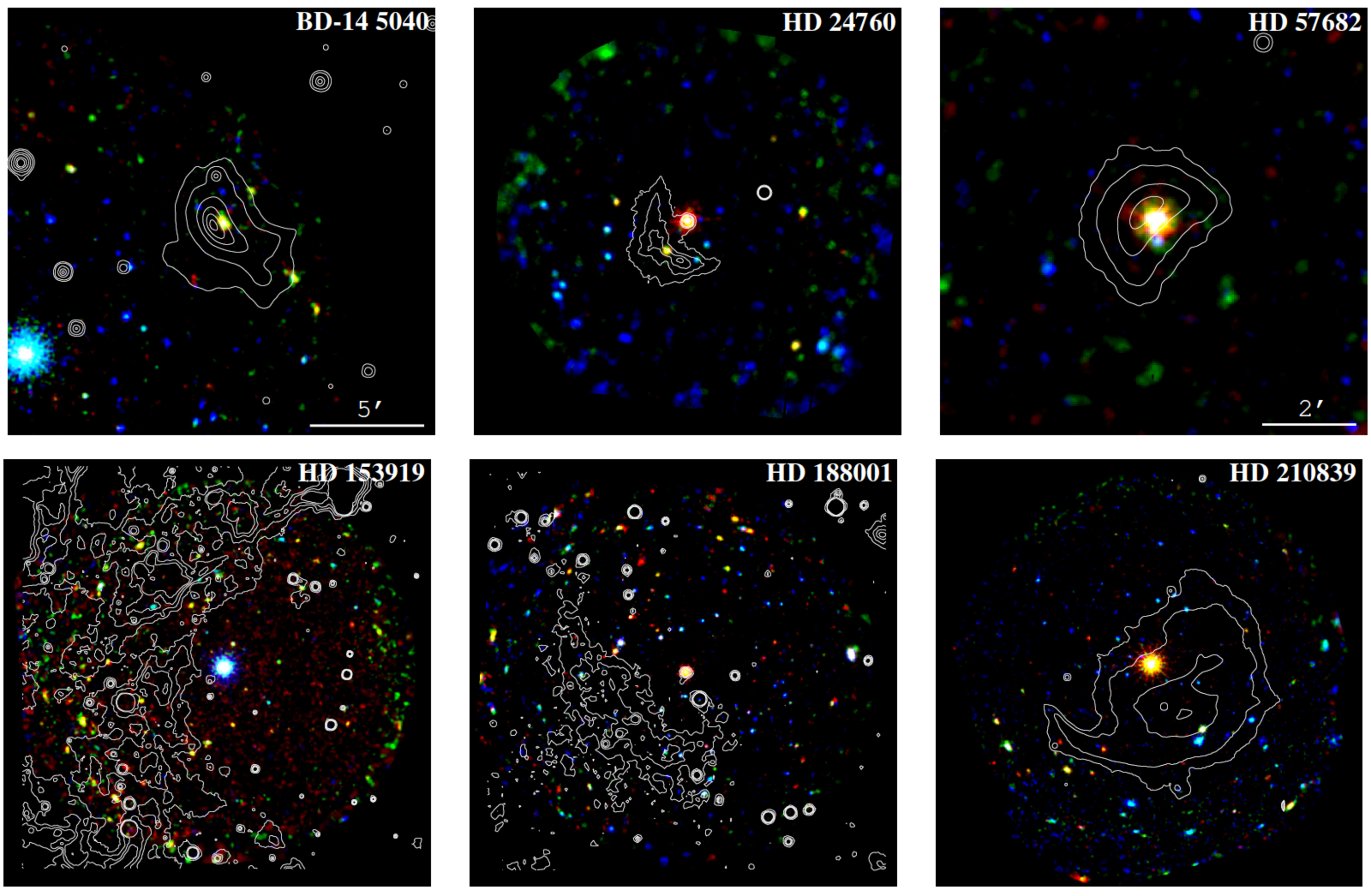}
\label{fig:WISE_RGB}
\caption{{\it XMM-Newton} EPIC (pn$+$MOS1$+$MOS2) exposure-corrected,
  background-subtracted X-ray images of the six bow shocks studied in
  this paper. Red, green, and blue correspond to the soft (0.3 -
  1.0~keV), medium (1.0 - 2.0~keV), and hard (2.0 - 5.0~keV) bands,
  respectively. The progenitor stars are placed at the center of each
  panel. Contours show the {\it WISE} W4 22~$\mu$m emission from the
  bow shock on each panel. North is up, east is to the left.}
\end{center}
\end{figure*}

\begin{table}
\caption{Estimated X-ray fluxes}
\begin{center}
\begin{tabular}{cccc}
\hline
\hline
Object                & $N_\mathrm{H}$                & $f_\mathrm{X}$           & $F_\mathrm{X}$ \\
                      & $\times$10$^{21}$~cm$^{-2}$ & erg~cm$^{-2}$~s$^{-1}$ &  erg~cm$^{-2}$~s$^{-1}$ \\
\hline
BD$-$14$^{\circ}$5040  & 10.7                        & 3.3$\times$10$^{-15}$    & 6.4$\times$10$^{-15}$ \\
HD\,24760             & 2.8                         & 1.0$\times$10$^{-14}$    & 1.4$\times$10$^{-14}$ \\
HD\,57682             & 4.4                         & 4.2$\times$10$^{-15}$    & 6.6$\times$10$^{-15}$ \\
HD\,153919            & 7.2                         & 7.0$\times$10$^{-15}$    & 1.2$\times$10$^{-14}$ \\
HD\,188001            & 2.8                         & 1.7$\times$10$^{-16}$    & 2.5$\times$10$^{-16}$ \\
HD\,210839            & 9.2                         & 7.4$\times$10$^{-15}$    & 1.4$\times$10$^{-14}$ \\
\hline
\hline
\end{tabular}
\end{center}
\end{table}

\section{Discussion}

The detection of non-thermal radio emission associated with the bow
shock around BD$+$43$^{\circ}$3654 \citep{Benaglia2010} opened a door
for studies of particle acceleration by massive stars. Those VLA
observations showed that the non-thermal emission has an extended
distribution, spatially coincident with the bow shock observed by the
Midcourse Space Experiment (MSX) in the D (14.65~$\mu$m)
band\footnote{We note that fig.~3 in \citet{Benaglia2010} shows a
  clump with positive spectral index which is not associated with the
  bow shock, in fact this feature is a clump easily spotted in the
  {\it WISE} W3 12~$\mu$m image presented by
  \citet{Toala2016}.}. Since then, the idea that charged particles in
the compress ISM (the bow shock) can cool down by non-thermal
processes (such as syncrotron radiation), has been studied extensively
in theoretical studies \citep[see][and references
  therein]{Pereira2016}. But the absence of firmly confirmed
detections of high-energy non-thermal emission is pushing the limits
of theory, even for the case of BD$+$43$^\circ$3654.

In order to estimate upper limits to the non-thermal diffuse X-ray
emission, we extracted background-subtracted spectra from regions
spatially coincident with the bow shocks in our sample. The obtained
background count rates in the 0.3--5.0~keV energy range along with the
estimated hydrogen column densities ($N_\mathrm{H}$)\footnote{We used
  the {\it Chandra} Galactic Neutral Hydrogen Density Calculator:
  \url{http://cxc.harvard.edu/toolkit/colden.jsp}} were used to obtain
absorbed ($f_\mathrm{X}$) and unabsorbed ($F_\mathrm{X}$) X-ray
fluxes. Using the {\it Chandra} PIMMS
tool\footnote{\url{http://cxc.harvard.edu/toolkit/pimms.jsp}} we
estimated the fluxes assuming that the emission can be modeled by a
power-law spectrum with $\Gamma=1.5$. Table~2 shows that our estimated
observed fluxes are comparable to those reported by \citet{Toala2016}
for $\zeta$\,Oph and BD$+$43$^{\circ}$3654.

Our systematic search using a clean sample of runaway massive stars
add to the list of bow shocks around runaway stars without non-thermal
diffuse X-ray emission \citep[][]{Terada2012,Toala2016}. Our present
results encouraged us to question the previously claimed detection of
non-thermal X-ray emission in the bow shock around AE~Aur
\citep{LopezSantiago2012}. In Appendix~A, we show that the detected
emission is a point-like source unrelated to the bow shock around
AE~Aur.

It has become evident that current theoretical models overpredict the
flux of the non-thermal diffuse X-ray emission in bow shocks around
runaray. To start with, \citet{delValle2012} adopted a mass-loss rate
a factor of $\sim$5 greater than that reported by \citet{Gva2012},
overestimating the density of high-energy particles. On the other
hand, unlike the cases of supernova remnants which are known to emit
considerably non-thermal X-ray emission \citep[e.g.,][]{Bamba2005},
the open morphologies of bow shocks around runaway stars might reduce
the injection efficiency of energy from thermal plasma to accelerate
particles and produce non-thermal emission.

\section{Conclusions}

We have searched for non-thermal diffuse X-ray emission associated
with bow shocks around runaway O-type stars. We used {\it XMM-Newton}
observations of a sample of 6 well determined runaway stars and found
no evidence of such emission.

We also revised the only claimed case of non-thermal diffuse X-ray
emission detected from a bow shock, AE~Aur. There is emission;
however, its spatial distribution is consistent with being that of a
point source.  Moreover, this X-ray source is not spatially coincident
with the bow shock. Thus, we conclude that this X-ray emission is not
associated with the bow shock.

Thus far, there are 9 bow shocks around O~stars that stand in defiance
of the recent and growing body of theoretical predictions for
non-thermal diffuse X-ray emissions from such structures. We conclude
that, if this predicted non-thermal diffuse X-ray emission is present
in bow shocks around runaway O stars, it is below the detection limits
of the current X-ray satellites.
\\

\begin{figure*}
\begin{center}
  \includegraphics[angle=0,width=.34\linewidth]{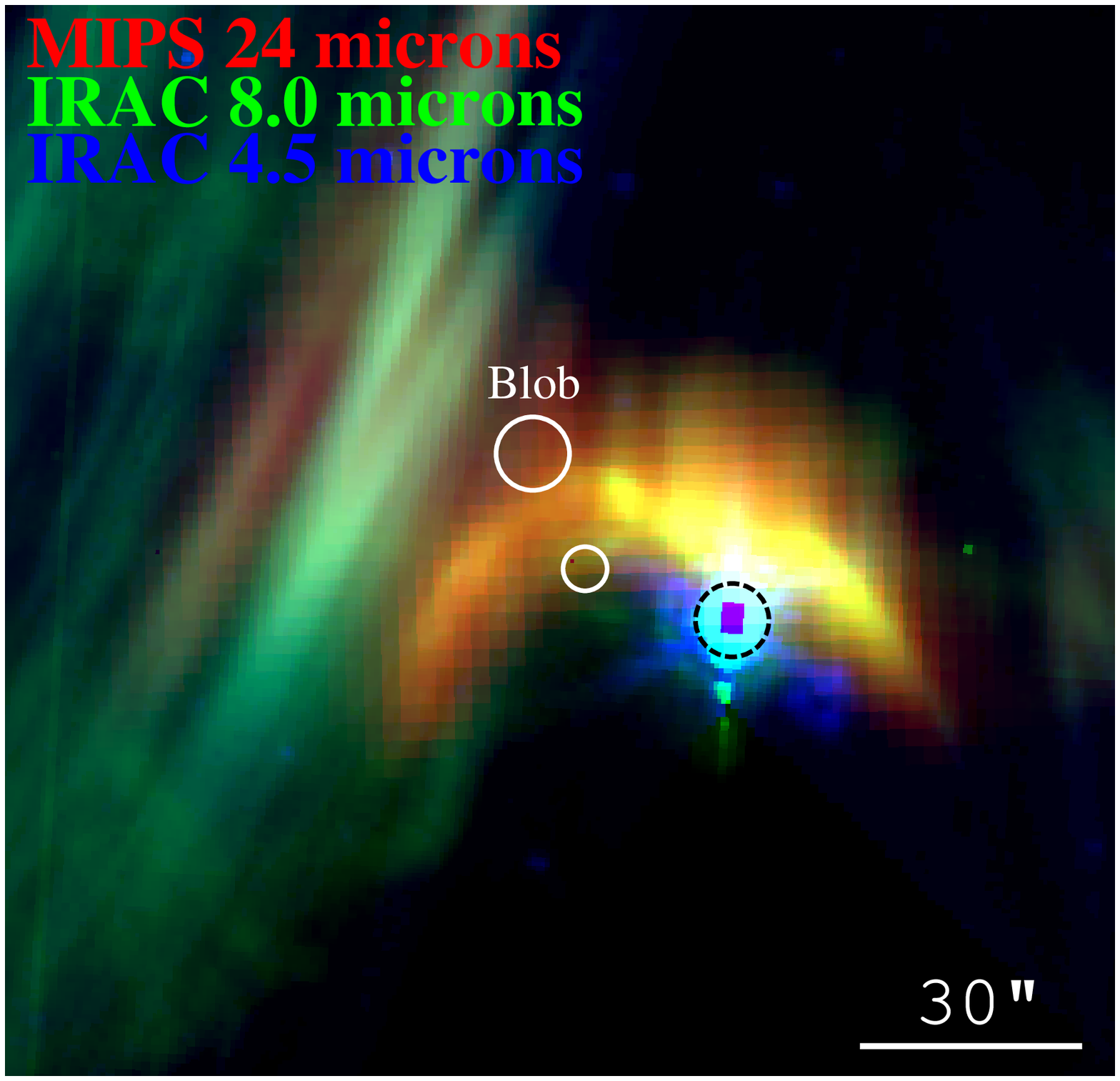}~
  \includegraphics[angle=0,width=.34\linewidth]{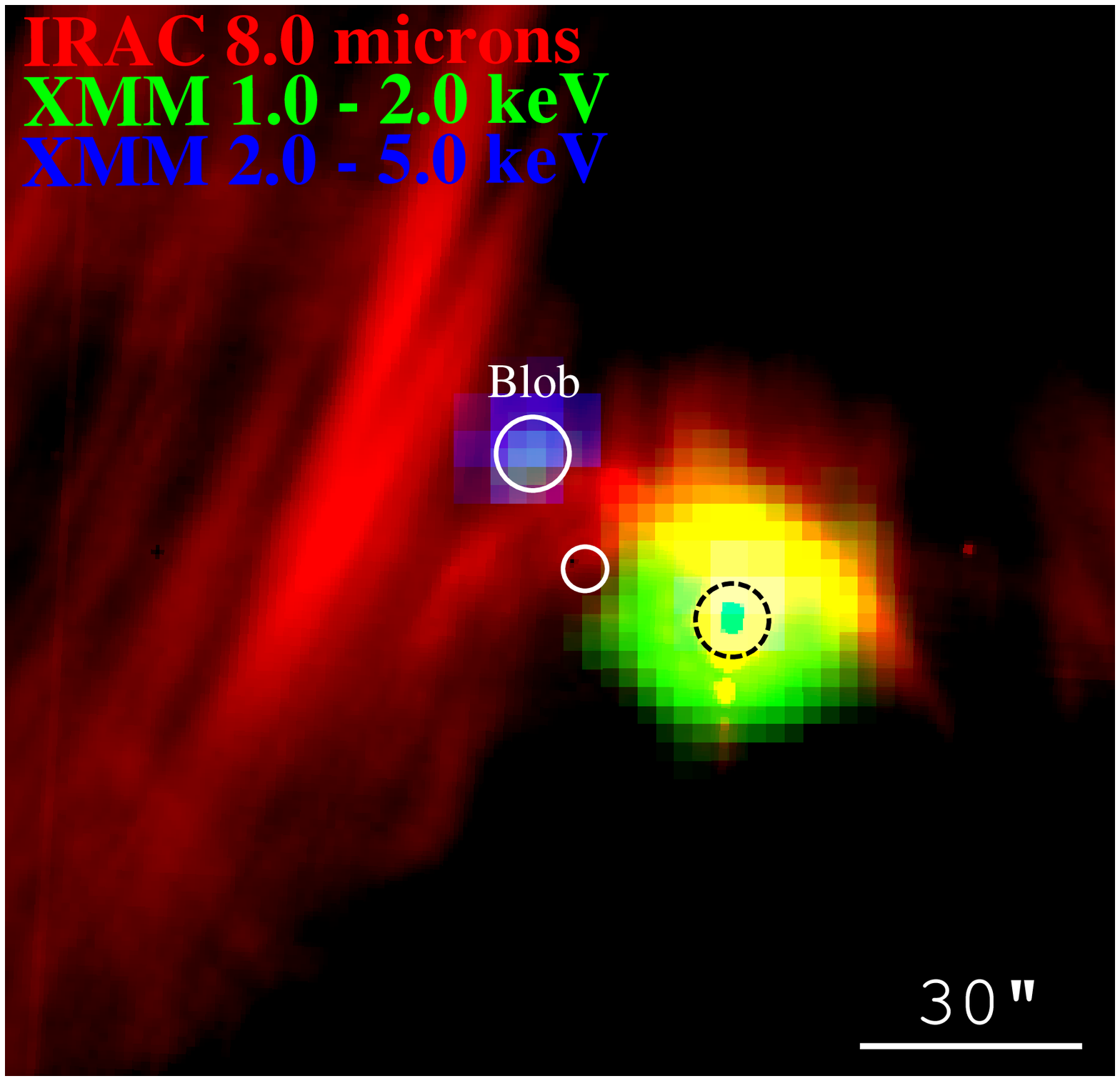}
\label{fig:WISE_RGB}
\caption{Composite colour images of the bow shock around AE
  Aurigae. North is up, and East is right.  Left: Red, green, and blue
  correspond to {\it Spitzer} MIPS 24~$\mu$m, IRAC 8~$\mu$m, and IRAC
  4.5~$\mu$m, respectively. Right: Red corresponds to the {\it
    Spitzer} MIPS 8~$\mu$m observation whilst green and blue
  correspond to the medium (1.0 - 2.0~keV) and hard (2.0 - 5.0~keV)
  bands, respectively. The position of AE~Aur is shown with a (black)
  dashed-line circular aperture and the position of the X-ray blob is
  shown with a (white) solid-line circular aperture whilst the
  position of the CO globule \#5 as reported by \citet{Gratier2014} is
  shown by the smaller circular aperture.}
\end{center}
\end{figure*}


The authors would like to thank the anonymous referee for valuable
comments that improved our manuscript. This work was based on
observations obtained with {\it XMM–Newton}, an ESA science mission
with instruments and contributions directly funded by ESA Member
States and NASA. This publication also makes use of data obtained with
{\it WISE} and {\it Spizer}. 
({\it WISE})


\appendix

\section{AE\,Aurigae}

The absence of non-thermal diffuse X-ray emission towards the sample
of six bow shocks presented in this paper, along with previously
reported undetections (see Section~1), questions the nature and
presence of the X-ray emission towards AE~Aur reported by
\citet{LopezSantiago2012}.  To confirm previous results, we have
analyzed the {\it XMM-Newton} observations of AE~Aur in a similar way
as described for other sources studied here. We compare our X-ray
images to {\it Spizer} MIPS and IRAC images. 

The left panel of Figure~3 shows the higher-resolution image of the
{\it Spitzer} data as compared to the {\it WISE} W3 image \citep[see
  figure 1 in][]{LopezSantiago2012}. On the other hand, the right
panel of Figure~3 presents a comparison of the {\it Spizer} IRAC
8~$\mu$m and the medium and hard X-ray bands. This panel confirms that
the blob of X-ray emission has a point-like shape with an angular
separation of 35\arcsec\, from AE~Aur, but also that this emission is
not spatially coincident with the bow shock around AE~Aur. The
position of the blob of X-ray emission is shown in Fig.~3-right panel
with a (white) solid line circular aperture. This X-ray blob is not
one of the dense molecular globules detected in CO $\sim$25\arcsec\,
from AE~Aur \citep[see globule \#5 in figure~2
  of][]{Gratier2014}. Thus, we can not confirm the previous claims on
non-thermal emission associated with the bow shock around AE~Aur.

\end{document}